\documentclass[twocolumn,english,aps,prd,reprint,floatfix,notitlepage,footinbib,preprintnumbers,superscriptaddress,altaffilletter]{revtex4-2}

\usepackage{amsmath,amssymb,amsfonts}
\usepackage{hyperref,breakurl,cleveref,url}
\usepackage{color}

\usepackage{graphicx}
\usepackage[export]{adjustbox}

\usepackage{accents,mathrsfs,mathtools}

\renewcommand{\paragraph}[1]{%
	\textit{#1}.---%
}

\def\skip{\vskip1.5pt}
\newcommand\trick[1]{}

\usepackage{enumitem}
\setlist[enumerate]{
	label={},
	leftmargin=2em,
	itemsep=2pt,
	topsep= 2pt,
	partopsep=0pt,
	parsep=0pt,
}

\let\oldeqref\eqref
\renewcommand{\eqref}[1]{Eq.\,\smash{\oldeqref{#1}}}
\newcommand{\eqrefs}[2]{Eqs.\,\smash{\oldeqref{#1}} and \smash{\oldeqref{#2}}}

\let\oldcite\cite
\renewcommand{\cite}[1]{\smash{\oldcite{#1}}}
\newcommand{\rcite}[1]{Ref.\,\smash{\oldcite{#1}}}
\newcommand{\rrcite}[1]{Refs.\,\smash{\oldcite{#1}}}

\newcommand{\fref}[1]{Fig.\,\smash{\ref{#1}}}

\def\JHK{{J.-H.\,K. }}

\def\Re{{\operatorname{Re}}}

\def\mem{\hspace{0.1em}}
\def\hem{\hspace{0.05em}}
\def\nem{\hspace{-0.1em}}
\def\hnem{\hspace{-0.05em}}
\def\hhem{\hspace{0.025em}}

\def\qiq{{\quad\implies\quad}}

\def\a{\alpha}
\def\b{\beta}
\def\c{{\gamma}}

\def\e{\epsilon}
\def\ve{\varepsilon}
\def\m{\mu}
\def\n{\nu}
\def\r{\rho}
\def\s{\sigma}
\def\k{\kappa}

\def\mdot{{\mem\cdot\mem}}

\def\da{{\dot{\a}}}
\def\db{{\dot{\b}}}
\def\dc{{\dot{\c}}}

\def\tdo{\tilde{o}}
\def\ti{\tilde{\iota}}

\newcommand{\wrap}[1]{{\smash{#1}\vphantom{\beta}}}

\def\lsq{{
		\kern-0.037em
		\adjustbox{scale=0.919,valign=c}{$
			{
				\adjustbox{raise=-0.0855em}{$\lfloor$}
				\llap{\reflectbox{\rotatebox[origin=c]{180}{$\lfloor$}}}
			}
			$}
		\kern-0.04em
}}
\def\rsq{{
		\kern-0.04em
		\adjustbox{scale=0.919,valign=c}{$
			{
				\rlap{\reflectbox{\rotatebox[origin=c]{180}{$\rfloor$}}} 
				\adjustbox{raise=-0.0855em}{$\rfloor$}
			}
			$}
		\kern-0.037em
}}

\def\vex{\vec{x}}
\def\vea{\vec{a}}

\def\tzeta{{\protect\tilde{\zeta}}}

\def\tm{{\tilde{m}}}

\def\tchi{\tilde{\chi}}

\def\tell{\smash{\tilde{\ell}}}
\def\trho{\smash{\tilde{\rho}}}
\def\ttheta{\smash{\tilde{\theta}}}
\def\teta{\smash{\tilde{\eta}}}

\def\robl{{r}}
\def\rprol{{\tilde{r}}}

\def\zag{{\includegraphics[valign=c]{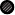}}}
\def\zig{{\includegraphics[valign=c]{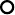}}}

\def\R{\mathbb{R}}
\def\C{\mathbb{C}}

\def\mflat{\mathbb{M}}

\newcommand{\BB}[1]{\Big(\,{#1}\,\Big)}

\begin{document}
	
	\title{
		Newman-Janis Algorithm from Taub-NUT Instantons
	}
	
	\author{Joon-Hwi Kim}
	\affiliation{Walter Burke Institute for Theoretical Physics, California Institute of Technology, Pasadena, CA 91125}
	
	\begin{abstract}
		It is shown that the Kerr metric represents the nonlinear superposition of self-dual and anti-self-dual Taub-NUT instantons.
		This promotes the Newman-Janis algorithm to a rigorous derivation of the Kerr metric with a definite physical origin.
		In the same way, the Kerr-Newman and charged Kerr-Taub-NUT solutions are systems of Taub-NUT instantons and chiral dyons.
	\end{abstract}
	
	\preprint{CALT-TH 2024-051}
	
	\bibliographystyle{utphys-modified}
	
	\renewcommand*{\bibfont}{\fontsize{8}{8.5}\selectfont}
	\setlength{\bibsep}{1pt}
	
	\maketitle
	
\paragraph{Introduction}%
The Newman-Janis algorithm (NJA) \cite{Newman:1965tw-janis} 
is
a mystery in 
the history of relativity.
Discovered
about sixty years ago,
the NJA
aims to
derive the
rotating black hole solution of Kerr \cite{Kerr:1963ud}
by applying a ``complex coordinate transformation''
to the 
static black hole solution of Schwarzschild.
Apparently, this procedure
is a sequence of ad hoc manipulations.
Each element in the Schwarzschild metric
must be
promoted to 
specific combinations of complex variables,
which are then transformed by a rule
of unknown physical origin.
Hence
the NJA has stood merely as
a formal mathematical trick.

Such a lack of clear understanding
is unsettling
because the NJA
holds the historical significane
as the very method that led to the discovery of
the Kerr-Newman metric
\cite{Newman:1965my-kerrmetric}:
the charged rotating black hole solution.

Despite early skepticism---%
epitomized by characterizations such as
a ``fluke'' \cite{Drake:1998gf}
or ``methods which transcend logic''  \cite{ernst1968new2}---%
research efforts have persisted over decades
toward a satisfactory justification
\cite{%
	%
	Erbin:2014aja,Erbin:2014aya,Erbin:2016lzq,Brauer:2014wwa,Keane:2014sta,Ferraro:2013oua,%
	%
	Whisker:2008kk,Erbin:2015pla,Erbin:2014lwa,Mirzaiyan:2017adt,Tavakoli:2020uzr,Yazadjiev:1999ce,CiriloLombardo:2004qw,Roberts:1988un,bambi2013rotating,%
	%
	Azreg-Ainou:2014pra,ernst1968new1,ernst1968new2,%
	%
	Rajan:2015ffs,Rajan:2016qiy,Rajan:2016zmq,giampieri1990introducing,quevedo1992complex,harvey1989complex,santos1975newtonian,flaherty1976hermitian,Aksteiner:2022bwr,%
	%
	Ayon-Beato:2015nvz,Lan:2024wfo,kamenshchik2023newman,beltracchi2021physical1,beltracchi2021physical2,Canonico:2011lba,canonico2010theoretical,%
	%
	Drake:1997hh,Drake:1997hh,herrera1982complexification,ibohal2005rotating,viaggiu2006interior,%
	%
	Talbot:1969bpa,schiffer1973kerr,Gurses:1975vu,Finkelstein:1974nr,demianski1972new,demianski1966combined,Carter:1968rr,Drake:1998gf,%
	Kerr:2007dk,Adamo:2014baa,Adamo:2009vu,%
	%
	newman1988remarkable,Newman:1973afx,Newman:2002mk,Newman:1973yu,newman1974curiosity,newman1974collection,newman1973complex,newman2004maxwell,%
	newman1976heaven,ko1981theory,grg207flaherty,%
	%
	sst-asym,ambikerr1,gmoov,%
	note-sdtn,%
	Crawley:2021auj,%
	gabriel1,%
	Guevara:2018wpp,Guevara:2019fsj,chkl2019,aho2020%
}.
These works may be classified into three categories, as outlined below.

The first category could be named
\textit{classical explanations}.
Here, the goal is to justify the NJA
from geometrical perspectives.
This means to
resolve inherent ambiguities,
identify hidden assumptions,
and explore simplifications or generalizations.
Notable contributions 
in this direction
include works by
Talbot \cite{Talbot:1969bpa},
Drake and Szekeres \cite{Drake:1998gf},
G\"urses and G\"ursey \cite{Gurses:1975vu},
Flaherty \cite{flaherty1976hermitian,grg207flaherty},
Rajan and Visser \cite{Rajan:2016zmq},
and
Giampieri \cite{giampieri1990introducing}.
The original article \cite{Newman:1965tw-janis}
also provides a brief justification
along this direction:
an allowed possibility in Kerr theorem \cite{penrose1967twistoralgebra}.
These appr\-oaches could be agnostic about physical origins, however.

The second category describes
\textit{the works of Newman himself}
\cite{%
	newman1988remarkable,Newman:1973afx,Newman:2002mk,Newman:1973yu,newman1974curiosity,newman1974collection,newman1973complex,newman2004maxwell,%
	newman1976heaven,ko1981theory,grg207flaherty%
},
which portray a more serious stance.
Emphasis is put on the fact that
the NJA realizes the spin of Kerr black hole
via an imaginary displacement into complexified spacetime.
The NJA is treated not as a mere mathematical curiosity
but a signal of a deeper physical structure,
such as
a Hodge duality
on angular momenta as
mass dipole moments
\cite{newman1974curiosity}.
A fundamental unification of spin and spacetime
is envisioned from
the construction of ``complex center of mass''
\cite{newman1974curiosity,newman1974collection,newman1973complex,newman2004maxwell,Newman:1973yu,Newman:2002mk,newman1976heaven,ko1981theory,grg207flaherty,sst-asym}.

Lastly, the third category
describes the \textit{modern explanation} emerged through works
\cite{ahh2017,Guevara:2018wpp,Guevara:2019fsj,chkl2019,aho2020}.
Here, one adopts
a particle physicist's perspective
and imagines observing the Kerr black hole
by throwing gravitons
(i.e., through gravitational waves)
from far away,
as if one detects the internal structure of a proton
by firing electrons 
in a particle collider.
In terms of scattering cross-sections,
it is revealed
that
the Kerr black hole
behaves like 
an ideally pointlike object
when the incident gravitons are all
circularly polarized
in either right or left handedness,
i.e.,
self-dual (SD) or anti-self-dual (ASD)
\cite{Guevara:2018wpp,Guevara:2019fsj,chkl2019,aho2020,Johansson:2019dnu,Aoude:2020onz,Lazopoulos:2021mna,zihan2023,fabian2}.
Thus,
despite the black hole's
extended geometrical features
such as the ring singularity or the horizon,
its ``graviton X-ray image''
displays just two objects:
a point that
absorbs SD gravitons only 
and another 
point
that
absorbs ASD gravitons only.
This stunning simplicity
is identified as the basis of NJA \cite{aho2020},
as
the coordinates of such points
precisely predict Newman's complex centers of mass
by taking complex values.

\begin{figure}[t]
	\centering
	\includegraphics[height=70pt,valign=c,
	clip= true,
	trim= 2.5pt 2.5pt 2.5pt 2.5pt
	]{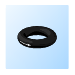}
	\quad
	\adjustbox{scale=1.5,valign=c}{$=$}
	\quad
	\includegraphics[height=70pt,valign=c,
	clip= true,
	trim= 2.5pt 2.5pt 2.5pt 2.5pt
	]{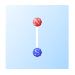}
	\caption{
		In electromagnetism,
		a current loop
		behaves like
		a pair of
		opposite
		monopoles:
		Amp\`erian and Gilbertian dipoles
		\cite{griffiths3ed}.
		This letter reveals
		a gravitational version of this duality.
	}
	\label{fig:gad}
\end{figure}

The modern explanation shows that
the NJA is not 
a mere mathematical construct
but rather a representation of some physical structure:
the NJA is recast as a statement about scattering amplitudes,
a gauge-invariant physical observable.
That said,
it did not decipher every aspect of the NJA
in its original, relativist form
as a statement about bulk metrics.
The conversion of the scattering amplitudes to bulk metrics
is based on
perturbative methods
\cite{Duff:1973zz,Neill:2013wsa,xanthopoulos1978exact,Harte:2016vwo,vines2018scattering},
so a resummed picture is lacking.

A similar caveat applies to the fact
attributed to
Crawley, Guevara, Miller, and Strominger \cite{Crawley:2021auj}
or
Ghezelbash, Mann, and Sorkin \cite{Ghezelbash:2007kw},
which is limited to the SD sector
and thus fall short of 
fully explaining the nonlinear aspect of NJA
with both SD and ASD components.

In this letter,
we provide 
a principled derivation of the Kerr metric
that faithfully reproduces the NJA
in its original form.
This not only identifies 
a definite physical origin
for structures observed in the classical explanations
and Newman's works
but also reveals a fully nonlinear, nonperturbative picture
completing the modern explanation.

The key new finding is that
the Kerr metric describes
a pair of SD and ASD Taub-Newman-Unti-Tamburino (NUT) instantons,
which are known to be gravitational analogs of chiral dyons.
These instantons
are exactly the ideally pointlike objects
in the field theorists' 
graviton X-ray image.
The Kerr metric is derived by
establishing a mathematical theorem
about nonlinear superposition of 
Kerr-Schild (KS) spacetimes.
This builds upon
a crucial recent finding \cite{note-sdtn}
and
a modern formalism
known as
KS double copy
\cite{monteiro2014black}.
Our construction generalizes to
charged Kerr-Taub-NUT solutions,
thus validating the historical derivation \cite{Newman:1965my-kerrmetric} of the Kerr-Newman metric
as well.

\skip
\paragraph{Intuition and Overview}%
The prototype of this paper 
grew out of
a moment of pure imagination 
that struck the present author
some years ago.
In electromagnetism,
a magnetic dipole arises from
not only
an Amp\`erian current loop
but also
a static monopole pair,
like the N and S poles of a bar magnet
(\fref{fig:gad}).
By analogy,
the question
was whether
the ring singularity of Kerr black hole
can be dualized into a pair of gravito-magnetic monopoles.

It is known
\cite{Misner:1963flatter,Bonnor:1969ala,sackfield1971physical,demianski1966combined,Plebanski:1975xfb,dowker1974nut,Griffiths:2009dfa}
that the gravitational analog of the monopole
is the Taub-NUT \cite{Taub:1950ez,Newman:1963yy} solution.
This
is a stationary vacuum solution characterized by two parameters:
mass $M$ and magnetic mass $N$,
which are gravito-electric and gravito-magnetic charges, respectively.
The metric of the Taub-NUT solution develops a semi-infinite line defect 
as the gravitational Dirac string,
dubbed Misner string
\cite{Misner:1963flatter,Bonnor:1969ala}.

A modern framework 
known as KS double copy \cite{monteiro2014black}
has elevated
such analogies
to exact mathematical correspondences
between
stationary vacuum solutions in Maxwell theory and general relativity.
For example,
the Coulomb solution
is the counterpart of
the Schwarzschild solution.
It has been recently shown \cite{note-sdtn} that
this correspondence applies to the Taub-NUT solution as well,
yet only in chiral (SD or ASD) limits
in which
$M = \pm i\mem N$.

The chiral limits of the Taub-NUT solution
are referred to as Taub-NUT instantons.
Historically, they were introduced
by the seminal works of Gibbons and Hawking
\cite{hawking1977gravitational,Gibbons:1978tef}
as paradigmatic instances of gravitational instantons;
hence the name.

A previously \textit{known fact}
is that 
any number of
Taub-NUT instantons of the same chirality
can be superposed together
to yield a multi-centered exact solution
\cite{hawking1977gravitational,Gibbons:1978tef}.
The \textit{new fact}
we prove
is that
SD and ASD Taub-NUT instantons can be superposed
to give a two-centered exact solution:
opposite chiralities.
Remarkably, 
this two-centered solution is nothing other than the Kerr solution.

\skip
\paragraph{Factorization of Kerr}%
As an introduction,
we begin with an innocuous mathematical fact
about a quartic polynomial.
Consider the zero locus
\begin{align}
	\label{eq:variety}
	(x^2{\,+\,}y^2{\,+\,}z^2 - a^2)^2
	+ (2az)^2 = 0
	\,,
\end{align}
where $a$ is a constant.
\eqref{eq:variety} describes a \textit{ring} of radius $a$
in the three-dimensional space $(x,y,z) \in \R^3$.
Intriguingly,
this ring can also appear as a set of
\textit{two disjoint points}.
Suppose we allow $x,y,z$ in \eqref{eq:variety}
to take complex values
so that
we obtain
a complex algebraic variety
in the affine space $\mathbb{C}^3$.
Then the ring is the slice of this variety
by the real section $x,y,z \in \mathbb{R}$.
However,
consider the slice
$x,y,z \in i\hem\mathbb{R}$.
Wick-rotating $x,y,z$
in \eqref{eq:variety} gives
\begin{align}
	\label{eq:variety-wick}
	(x^2{\,+\,}y^2{\,+\,}z^2 + a^2)^2
	- (2az)^2 = 0
	\,,
\end{align}
which factorizes to
$(x^2 + y^2 + (z+a)^2)(x^2 + y^2 + (z-a)^2) = 0$.
This admits just two points as solutions:
$(0,0,\pm a)$.

The astute reader will point out that
\eqref{eq:variety}
is the very equation of the ring singularity
of the Kerr black hole.
The Kerr metric is characterized by 
mass $M$ and ring radius $a$.
In KS coordinates, for instance, it reads
\begin{align}
	\label{eq:line-in-real}
	&
	{-dt^2} {\,+\,} dx^2 {\,+\,} dy^2 {\,+\,} dz^2
	\\
	&{
		+\mem
		\frac{2M \robl^3}{\robl^4{\,+\,}a^2z^2}
		\bigg({
			-dt 
			+ \frac{\robl x{\,-\,}ay}{\robl^2{\,+\,}a^2}\mem dx
			+ \frac{\robl y{\,+\,}ax}{\robl^2{\,+\,}a^2}\mem dy
			+ \frac{z}{\robl}\mem dz
		}\mem\bigg)^{\hnem\nem2}
	}
	\,.
	\nonumber
\end{align}
Here, $\robl$ is a function of $x,y,z$
implicitly defined by
\begin{align}
	\label{eq:r-oblate}
	\frac{x^2{\,+\,}y^2}{\robl^2{\,+\,}a^2}
	+ \frac{z^2}{\robl^2}
	= 1
	\,.
\end{align}
The contour surfaces of $\robl$
are oblate spheroids,
which degenerate to a disc 
of radius $a$
at $\robl = 0$.
The boundary of this disc
is exactly the ring singularity
shown in \eqref{eq:variety}.

\begin{figure}[t]
	\centering
	\adjustbox{valign=c}{
		\includegraphics[scale=0.9
		,clip=true,trim=7pt 7pt 7pt 7pt
		]{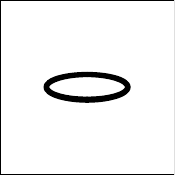}
	}
	\,\,$\xrightarrow[]{\,\,\textsc{\scriptsize{Wick}}\,\,}$\,\,
	\adjustbox{valign=c}{
		\includegraphics[scale=0.9
		,clip=true,trim=7pt 7pt 7pt 7pt
		]{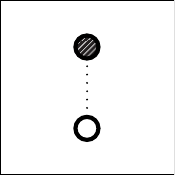}    
	}
	\caption{
		The ring singularity of Kerr black hole
		factorizes into
		a pair of
		SD and ASD
		Taub-NUT instantons.
	}
	\label{fig:factorization}
\end{figure}

By analytic continuation,
we regard
\eqref{eq:line-in-real}
as a holomorphic metric
on a complex four-manifold
solving
complexified vacuum Einstein's equations
(cf.\;\rcite{plebanski1975some}).
Then the total Wick rotation,
$x^\m \mapsto ix^\m$,
is an allowed coordinate change.
The resulting line element is
\begin{align}
	\label{eq:line-in-wick}
	&
	{dt^2} {\,-\,} dx^2 {\,-\,} dy^2 {\,-\,} dz^2
	\\
	&{
		+\mem
		\frac{2iM \rprol^3}{\rprol^4{\,-\,}a^2z^2}
		\bigg(\mem{
			dt 
			- \frac{\rprol x{\,+\,}iay}{\rprol^2{\,-\,}a^2}\mem dx
			- \frac{\rprol y{\,-\,}iax}{\rprol^2{\,-\,}a^2}\mem dy
			- \frac{z}{\rprol}\mem dz
		}\mem\bigg)^{\hnem\nem2}
	}
	\,,
	\nonumber
\end{align}
where $\rprol$ is
a function of 
the transformed
$x,y,z$:
\begin{align}
	\label{eq:r-prolate}
	\frac{x^2{\,+\,}y^2}{\rprol^2{\,-\,}a^2}
	+ \frac{z^2}{\rprol^2}
	= 1
	\,.
\end{align}
The level sets of $\rprol$ are
now prolate spheroids.
The line element in \eqref{eq:line-in-wick}
is ill-defined
over a ``needle'' along
$x,y=0$ and $z \in [-a,a]$,
which is the degenerate prolate spheroid at $\rprol = a$.
Curvature invariants imply that 
the boundary of this needle is only truly singular,
which exactly describes the set of two points 
specified by \eqref{eq:variety-wick}.

Direct computation shows that
these point singularities are
Taub-NUT instantons of opposite chiralities.
By taking $z \mapsto z \pm a$ and sending $a \to \infty$,
one finds that
the metric in the vicinity of the upper/lower tip of the needle
is exactly the SD/ASD Taub-NUT metric of \rcite{note-sdtn}.

Direct computation also shows that
the needle is a Misner string
that transports the magnetic mass flux
from one instanton to another.
The metric near any point on the needle
can be represented as
$
(\hhem
dt {\,-\,} 2iM\, d\log (x{\,-\,}iy)
)^2 $ $
- dx^2 - dy^2 - dz^2
$
\footnote{
	The line element diverges
	when zooming to an arbitrary point
	$z {\,=\,} \eta\hem a \in (-a,a)$.
	However, it
	can be compensated by a diffeomorphism 
	$
	(x{\,+\,}iy)
	\mapsto 
	(x{\,+\,}iy)
	-
	2iM
	(
	(1{\hem-\mem}\eta^2)\hem a {\,-\,} 2iM {\,-\,} 2\eta z
	)/(x{\,-\,}iy)
	$,
	$
	(x{\,-\,}iy)
	\mapsto
	(x{\,-\,}iy)
	$.
},
which
describes the image of
flat spacetime
under a large diffeomorphism.
Such a line defect of time monodromy
is the very notion of Misner string as a gravito-magnetic flux tube
\cite{dowker1967gravitational,deser1984three,Griffiths:2009dfa,misner1967taub,Alfonsi:2020lub,note-sdtn}.

In sum, we have explicitly shown that
the Kerr metric, as a \textit{whole},
represents a pair of SD and ASD Taub-NUT instantons.
In \fref{fig:factorization},
the instantons are visualized as blobs
while the Misner string is drawn as a dotted line:
\begin{align}
	\label{story1}
	\text{Kerr}
	\,\,\,\,=\,\,\,\,
	\zag 
	\:\:+\:\:
	\zig
	\,.
\end{align}

It is easily verified that their
masses, magnetic masses, and center positions
are
given as below
(if measured with respect to the slice of mostly-positive metric signature):
\begin{align}
	\label{eq:kerr-summary}
	\kern-0.3em
	\bigg\{\,
	\begin{aligned}
		\text{%
			SD instanton,
			\zag
		}
		\,&:\,
		(M/2,-iM/2)
		\,\,\,\text{at}\,\,\,
		\vex = +i\vea
		\,,\\
		\text{%
			ASD instanton,
			\zig
		}
		\,&:\,
		(M/2,+iM/2)
		\,\,\,\text{at}\,\,\,
		\vex = -i\vea
		\,.
	\end{aligned}
	\kern-0.1em
\end{align}
Note that the magnetic mass dipole moment computes to
{$\smash{\vec{J}} = (-iM/2)(2i\vea) = M\vea$},
which is precisely the spin angular momentum of the Kerr black hole.
In fact,
the static configuration in \eqref{eq:kerr-summary} reproduces
all the
black hole's
spin-induced 
mass multipole moments,
which take the same magnitude with alternating signs
\cite{janis1965structure,Newman:1965tw-janis,Newman:1973yu}.

The separation $2i\vea$ has to be pure-imaginary
for $\smash{\vec{J}}$ to be real,
since Taub-NUT instantons carry imaginary magnetic mass in Lorentzian signature.
In fact, 
we will learn that
the imaginary coordinates $\pm i\vea$
are exactly the imaginary displacements of the NJA.
Therefore,
we realize that
the NJA is a statement about
the curvature singularity of Kerr solution
when taken
as a complex saddle.

It should be stressed that our analysis above
does not count on any methods of linearization or perturbative expansions.
It is readily performed in other coordinate systems as well,
such as the Boyer-Lindquist system.

To our knowledge,
the above explicit proof of the exact equivalence between
Kerr and a dipolar configuration of Taub-NUT instantons
is new,
although
an argument appears in
\rcite{Gross:1983hb}.
Our result is also disparate from
the ``black dihole'' of
\rrcite{%
	Manko:2009xx,Clement:2018ynm,Clement:2021ukb,%
	bonnor1966exact,Emparan:1999au,Liang:2001sp,Emparan:2001bb,Teo:2003ug,Manko:2009zza,Chen:2012dr,Manko:2013iva,Cabrera-Munguia:2015fha,Manko:2017avt%
}.

\paragraph{Nonlinear Superposition Theorem}%
Next, we regard the Taub-NUT instantons as \textit{parts}
and show that
they can be assembled together
to form
the Kerr black hole:
\begin{align}
	\label{story2}
	\zag 
	\:\:+\:\:
	\zig
	\,\,\,\,=\,\,\,\,
	\text{Kerr}
	\,.
\end{align}
A crucial feature of Einstein's gravity theory,
however,
is nonlinearity of field equations.
Hence generally speaking,
it is virtually impossible to assemble two vacuum solutions together to obtain another
by a simple process.
The $+$ symbol in \eqref{story2}
signifies a \textit{nonlinear} sum.

Yet remarkably,
the Taub-NUT instanton solution
admits a KS metric \cite{note-sdtn}
as mentioned earlier.
For instance, the KS metric of a SD Taub-NUT instanton
is given in an elegant formula in the spinor notation:
\begin{align}
	\label{eq:tnks}
	ds^2 \,=\, ds^2_\text{flat} + 
	\frac{M}{|\vex|}\mem \Big(\,{
		\eta_\a\hhem \tdo_\da\mem dx^{\da\a}
	}\mem\Big)^2
	\,.
\end{align}
Here, the ASD spinor
$\eta_\a$ is an arbitrary constant reference
that controls the direction of the Misner string
(as an ASD null plane),
whereas
the SD spinor $\tdo_\da$
describes a ``square root'' of the position three-vector $\vex$
\footnote{
	Concretely, $\protect\tdo^{\protect\da}$ is the eigenspinor of 
	the Killing spinor
	\smash{$\protect\tchi^{\protect\da}{}_{\protect\db}  = \frac{1}{2}\mem (\protect\vex\mdot\protect\vec{\sigma})^{\protect\da}{}_{\protect\db}$}
	whose eigenvalue is $+|\protect\vex|/2$.
}.
That is,
the gauge and physical data
are respectively stored in the ASD and SD spinors.
The Weyl tensor is purely SD.
See \rcite{note-sdtn}
for the derivation of
\eqref{eq:tnks}
from the well-known
Gibbons-Hawking metric
\cite{hawking1977gravitational,Gibbons:1978tef}.

Notably,
KS metrics
are known to
enjoy an \textit{effective linearization} property:
the exact vacuum Einstein's equations
are solved if
the linearized equations
are solved
\cite{xanthopoulos1978exact,Harte:2016vwo,vines2018scattering}.
Via this effective linearization,
two vacuum KS metrics
can be assembled
into another
as shown below.

First of all,
a null vector field (NVF) $\ell$ in complexified Minkowski space is \textit{canonical}
if it is stationary, geodesic, shear-free, and unit-normalized as $\ell^0 = 1$
with respect to the time direction.
One can treat
canonical NVFs as projective objects,
due to the fixed normalization $\ell^0 = 1$.

Each canonical NVF
is associated with two solutions.
The \textit{single copy solution} is the electromagnetic gauge potential $A_\m = \theta\mem \ell_\m$.
The \textit{double copy solution} is the metric $g_{\m\n} = \eta_{\m\n} + \e\, \theta\mem \ell_\m \ell_\n$,
where $\e$ is an auxiliary parameter.
Provided that $\theta = \partial_\m \ell^\m$ is the expansion of the congruence of $\ell$,
$A_\m$ solves vacuum Maxwell's equations in flat spacetime
while $g_{\m\n}$ solves vacuum Einstein's equations.

The above set of definitions is
our formulation of the framework known as KS double copy \cite{monteiro2014black}.

Next, we present what we call
the \textit{superposition lemma}.
Suppose two canonical NVFs $\ell_1$ and $\ell_2$
with expansions $\theta_1$ and $\theta_2$.
Then
\begin{align}
	\label{eq:elformula-spinor}
	\ell^{\da\a}
	\,\propto\,\mem
	\ell_1^{\da\b} \delta_{\b\db}\mem \ell_2^{\db\a}
	\,,\quad
	\tell^{\da\a}
	\,\propto\,\mem
	\ell_2^{\da\b} \delta_{\b\db}\mem \ell_1^{\db\a}
\end{align}
are canonical NVFs
with expansions $\theta$ and $\tilde{\theta}$
such that
\begin{align}
	\label{eq:linearized-1}
	\theta_1\mem \ell_1{}_\m + \theta_2\mem \ell_2{}_\m
	\,&=\, \theta\mem \ell_\m + \tilde{\theta}\mem \tell_\m
	+ \partial_\m \chi
	\,,\\
	\label{eq:linearized-2}
	\theta_1\mem \ell_1{}_\m \ell_1{}_\n + \theta_2\mem \ell_2{}_\m \ell_2{}_\n
	\,&=\, \theta\mem \ell_\m \ell_\n + \tilde{\theta}\mem \tell_\m \tell_\n
	+ 2\mem \partial_\wrap{(\m} \xi_\wrap{\n)}
	\,,
\end{align}
and $\theta_1 + \theta_2 = \theta + \tilde{\theta}$.
Here,
$\delta_{\b\db} = (\s^0)_{\b\db}$
is the stationary direction boiled down into the spinor notation.
$\chi$ and $\xi_\n$ are some scalar and one-form fields.

If $\tilde{\theta}$ happens to vanish,
\eqref{eq:linearized-1} will imply that
the single copy solution of $\ell$
describes the superposition of the single copy solutions of $\ell_1$ and $\ell_2$.
Similarly,
\eqref{eq:linearized-2} will imply that
the double copy solution of $\ell$
describes the nonlinear superposition of the double copy solutions of $\ell_1$ and $\ell_2$.
First,
the linear sum $\theta_1\mem \ell_1{}_\m \ell_1{}_\n + \theta_2\mem \ell_2{}_\m \ell_2{}_\n$
is gauge-equivalent to $\theta\mem \ell_\m \ell_\n$
in linearized gravity.
Second, $\theta\mem \ell_\m \ell_\n$ uplifts to an exact solution
by the effective linearization property \cite{xanthopoulos1978exact,Harte:2016vwo,vines2018scattering} of KS metrics.

The \textit{nonlinear superposition theorem} is the following:
if two canonical NVFs $\ell_1$ and $\ell_2$ 
are respectively associated with SD and ASD single copy solutions,
then
the double copy solution of 
$\ell$
in \eqref{eq:elformula-spinor}
describes the nonlinear superposition of the double copy solutions of $\ell_1$ and $\ell_2$.
This is because $\tilde{\theta}$ vanishes.

The proofs are detailed in the supplemental material,
but we shall remark here that
the chiral nature of \eqref{eq:elformula-spinor} plays an important role:
the SD (ASD) spinor index $\da$ ($\a$) is controlled by the SD (ASD) solution's NVF $\ell_1$ ($\ell_2$).
Also, \eqref{eq:elformula-spinor}
translates into the vector notation as
\begin{align}
	\label{eq:elformula}
	\ell^\m
	\,=\,
	\frac{\mem{
			\ell_1^\m {\,+\,} \ell_2^\m 
			\mem+\mem 
			\ell_1\nem\mdot\ell_2\mem u^\m
			\mem+\mem
			i\mem \ve^\m{}_{\n\r\s}\mem u^\n\hem \ell_1^\r\mem \ell_2^\s
		}\mem}{2+\ell_1\mdot\ell_2}
	\,,
\end{align}
which describes
a highly nonlinear formula.

\skip
\paragraph{Derivation of Kerr Metric}%
The nonlinear superposition theorem provides
a \textit{derivation} of Kerr metric
(without quotation marks).

First,
note that
the SD Taub-NUT instanton metric in \eqref{eq:tnks}
is the double copy solution
of the canonical NVF
$\ell^{\da\a} \propto \tdo^\da\hem \eta^\a$,
whose single copy solution is SD
\cite{note-sdtn}.

Second,
prepare
a SD Taub-NUT instanton centered at $\vex_1$
and an ASD Taub-NUT instanton centered at $\vex_2$.
Their canonical NVFs are
$\ell_1^{\da\a} \propto \tdo_1^\da\mem \eta^\a$
and $\ell_2^{\da\a} \propto \teta^\da\mem o_2^\a$,
where
$\tdo^\da_1(\vex) = \tdo^\da(\vex{\,-\,}\vex_1)$
and
$o^\a_2(\vex) = o^\a(\vex{\,-\,}\vex_2)$
are ``square roots'' of
$\vex{\,-\,}\vex_{1,2}$.
The expansions are
$\theta_{1,2} = |\vex{\,-\,}\vex_{1,2}|^{-1}$,
as should be clear from \eqref{eq:tnks}.

Third,
construct their nonlinear superposition
via the nonlinear superposition theorem.
\eqref{eq:elformula-spinor} gives
\begin{align}
	\label{eq:o1o2}
	\ell^{\da\a}
	\,\,\propto\,\,\mem
	\tdo^\da_1\, \eta^\b
	\delta_{\smash{\b\db}\vphantom{\b}}\mem 
	\teta^\db\mem o^\a_2
	\,\,\propto\,\,\mem
	\tdo^\da_1\mem o^\a_2
	\,,
\end{align}
which simply transvects
the physical spinors 
$\tdo^\da_1$
and
$o^\a_2$.
The expansion is $\theta = \theta_1 + \theta_2
= |\vex{\,-\,}\vex_1|^{-1} + |\vex{\,-\,}\vex_2|^{-1}
$.

Finally,
the new vacuum spacetime
is obtained as
\begin{align}
	\label{eq:ig-kerr}
	\partial^2_\text{flat}
	- M\mem
	\bigg({
		\frac{1}{|\vex{\,-\,}\vex_1|}
		{\,+\,} 
		\frac{1}{|\vex{\,-\,}\vex_2|}
	}\bigg)
	\bigg(\mem{
		\frac{\tdo^\da_1 o^\a_2}{ 
			o_2^\c\mem \delta_{\c\dc}\mem \tdo_1^\dc
		}\, \partial_{\a\da}
	}\mem\bigg)^{\hnem\nem2}
	\,,
\end{align}
where we have written down the inverse metric
for a direct comparison with \rcite{Newman:1965tw-janis}
while setting $\e = M$.

Remarkably, the vacuum solution in \eqref{eq:ig-kerr}
is exactly the Kerr solution
with mass $M$ and angular momentum \smash{$\vec{J} = -iM\mem (\vex_1{\,-\,}\vex_2)/2$},
centered around $(\vex_1{\,+\,}\vex_2)/2$.

In particular, take 
$\vex_1 = +i\vea$ and $\vex_2 = -i\vea$ for $\vea = (0,0,a)$.
Then \eqref{eq:ig-kerr}
exactly coincides with the Kerr inverse metric
obtained in the very article \cite{Newman:1965tw-janis}.

This establishes a mathematically rigorous derivation of the Kerr solution
as the nonlinear superposition of SD and ASD Taub-NUT instantons.

\begin{figure}[t]
	{
		\centering
		\adjustbox{valign=c}{
			\includegraphics[scale=0.9
			,clip=true,trim=5pt 0pt 2pt 0pt
			]{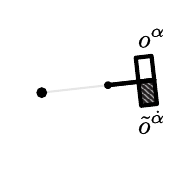}
		}
		$\xrightarrow[]{\textsc{\scriptsize{separate}}}$
		\adjustbox{valign=c}{
			\includegraphics[scale=0.9
			,clip=true,trim=5pt 0pt 2pt 0pt
			]{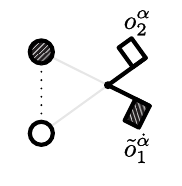}    
		}
	}
	\vspace{-1.0\baselineskip}
	\caption{
		In an accurate sense,
		the NJA
		first decomposed a Schwarzschild black hole
		into SD\,(\zag)
		and ASD\,(\zig) 
		Taub-NUT instantons
		and then moved them independently.
	}
	\label{fig:NJA}
\end{figure}

\skip
\paragraph{NJA Deciphered}%
It remains to explicate that
the above derivation faithfully reproduces the NJA.
Consider the inverse metric of the Schwarzschild solution,
identified as the nonlinear superposition of
SD and ASD Taub-NUT instantons overlapped at the same point:
\begin{align}
	\label{eq:ig-sch}
	\partial^2_\text{flat}
	- M\mem
	\bigg({
		\frac{1}{|\vex|}
		{\,+\,} 
		\frac{1}{|\vex|}
	}\bigg)
	\bigg(\mem{
		\frac{\hem \tdo^\da o^\a}{
			o^\c\mem \delta_{\c\dc}\mem \tdo^\dc
		}\, \partial_{\a\da}
	}\mem\bigg)^{\hnem\nem2}
	\,.
\end{align}
Each term here acquires a unique meaning
as per the nonlinear superposition theorem,
so
there exists a unique way of
deforming \eqref{eq:ig-sch}
such that the instantons get displaced to 
$\vex_1 = +i\vea$ and $\vex_2 = -i\vea$.
\eqref{eq:o1o2}
dictates the transformation of the spinors:
the SD (ASD) spinor shifts by $+i\vea$ ($-i\vea$).
The KS potential is mandated to be the total expansion:
$\theta_1 + \theta_2
= |\vex{\,-\,}i\vea|^{-1} + |\vex{\,+\,}i\vea|^{-1}
$.
Thus,
one
unambiguously
generates 
the Kerr inverse metric in
\eqref{eq:ig-kerr}
from 
the Schwarzschild inverse metric in
\eqref{eq:ig-sch}.
See \fref{fig:NJA} for a visualization.

This operation precisely reproduces the NJA in its very original formulation \cite{Newman:1965tw-janis},
as elaborated in the supplemental material.
In particular,
Eqs.\:(4) and (7)
of \rcite{Newman:1965tw-janis}
are \textit{literally} reproduced
upon using spheroidal coordinates.

In \rcite{Newman:1965tw-janis},
the $\pm i\vea$ imaginary shifts of the spinors (null tetrad)
and the ``complexification'' of the KS potential 
$2/|\vex| \to 2\mem\Re\mem |\vex{\,-\,}i\vea|^{-1}$
are posited as ad hoc rules.
The nonlinear superposition theorem
shows that this set of replacements, i.e., the NJA, is
a strict implication of the fact that
the Schwarzschild-Kerr family of solutions
are systems of SD and ASD Taub-NUT instantons.

This refined understanding implies the following
clarifications on the NJA.

First,
the NJA is \textit{not} a coordinate transformation
as originally conceived 
\cite{Newman:1965tw-janis}.
The instanton pairs with zero and nonzero separati\-ons,
i.e., Schwarzschild and Kerr,
are never diffeomorphic.

Second,
the NJA is not about one object
as originally conceived
\cite{Newman:1973yu,Newman:2002mk},
but \textit{two}.
It does not merely shifts
a Schwarzschild black hole
but instead \textit{splits} it into SD and ASD parts
and then \textit{separates} them,
as is illustrated in \fref{fig:NJA}.
The displacements $\vex_1, \vex_2$ are also \textit{independent},
although a reality condition can be imposed a posteriori.

Third,
it is important to explicitly address the nonlinear gravitational interaction
between the SD and ASD instantons.
Otherwise the NJA is not fully explained nor justified
as a method that generates an exact solution in Einstein gravity.
Newman's complex center of mass analysis
\cite{newman1974curiosity,Newman:1973yu,Newman:2002mk},
for instance,
identifies one worldline
from the SD sector (in linearized gravity)
and does not elaborate much on the validity of
``plus complex conjugate''
in the nonlinear theory.
Works \cite{Ghezelbash:2007kw,Crawley:2021auj}
have shown that
the SD Kerr-Taub-NUT solution
describes a SD Taub-NUT instanton,
which is not a statement about the Kerr solution which the NJA concerns.
To reiterate,
the NJA is about \textit{two} Taub-NUT instantons,
not just one SD instanton.

The technical structure of the formula in
\eqref{eq:elformula}
may have been effectively observed in
works \cite{schiffer1973kerr,Gurses:1975vu,Rajan:2016zmq},
but the instanton pair interpretation is not realized there.

\skip
\paragraph{NJA for Charged Kerr-Taub-NUT}%
Finally, we construct more general black holes.
First of all,
the electric-magnetic dual version of Kerr solution
arises as
\begin{align}
	\label{story2*}
	\:\:-\:\:
	\zag 
	\:\:+\:\:
	\zig
	\,\,\,\,=\,\,\,\,
	\text{Kerr}^\star
	\,.
\end{align}
The nonlinear superposition theorem
admits no freedom for 
relative coefficients,
but there a exist binary choice for the NVF
per each instanton.
In particular,
the KS metric of SD instanton
can be realized in either
outgoing ($\ell^{\da\a} \propto \tdo^\da \eta^\a$) or ingoing ($\ell^{\da\a} \propto  \ti^\da \eta^\a$)
conventions.
Switching to ingoing flips the sign of the expansion,
and in this sense
the instantons can be subtracted as in \eqref{story2*}.

\begin{figure}
	\adjustbox{valign=c}{
		\includegraphics[scale=0.9]{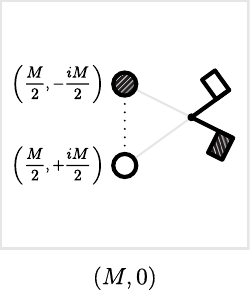}
	}
	\,
	\adjustbox{valign=c}{
		\includegraphics[scale=0.9]{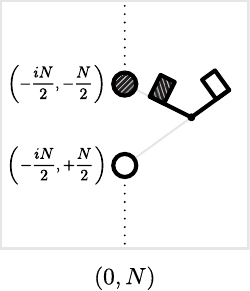} 
	}
	\caption{
		Implementing pure
		mass
		(\textit{left})
		and magnetic mass
		(\textit{right})
		with Taub-NUT instantons.
	}
	\label{fig:MN}
\end{figure}

Explicitly, the metric for
$\ell^{\da\a} \propto \ti^\da_1\hem o^\a_2$ is
\begin{align}
	\label{eq:Kerr*}
	&
	{dt^2} {\,-\,} dx^2 {\,-\,} dy^2 {\,-\,} dz^2
	\\
	&{
		-
		\frac{2 N a\rprol z}{\rprol^4{\hem-\,}a^2z^2}
		\bigg(\mem{
			dt 
			{\,+\,} \frac{\rprol}{a}
			\bigg({
				\frac{zx{\,+\,}i\rprol y}{\rprol^2{-\,}z^2}\mem dx
				{\,+\,} \frac{zy{\,-\,}i\rprol x}{\rprol^2{-\,}z^2}\mem dy
				{\,-\,} dz
			}\bigg)\nem\nem
		}\mem\bigg)^{\hnem\nem2}
	}
	\,,
	\nonumber
\end{align}
which describes 
a massless black hole with 
magnetic mass $N$
and gravito-electric dipole moment ${-\hnem N a}$.
The exercise
of zooming to each part of this metric
can be repeated
to explicitly verify the nonlinear superposition;
see \fref{fig:MN}.

By adding up the two KS perturbations in 
\eqrefs{eq:line-in-wick}{eq:Kerr*},
one obtains the generic three-parameter Kerr-Taub-NUT solution
in the double KS form
\cite{Plebanski:1975xfb,Chong:2004hw,Farnsworth:2023mff,Luna:2015paa}
\footnote{
	In the SD limit, $N \to -iM$,
	this double KS metric
	describes a SD Taub-NUT instanton,
	consistently with \rrcite{Ghezelbash:2007kw,Crawley:2021auj}.
}.
The two canonical NVFs
$\tdo^\da_1\hem o^\a_2$
and
$\ti^\da_1\hem o^\a_2$
share a common spinor $o^\a_2$,
making them mutually orthogonal.

Meanwhile,
the Maxwell stress-energy tensor
$\smash{T^{\da\a\db\b}} = \smash{\tilde{F}^{\da\db} F^{\a\b}}$
\cite{penrose1984spinors1}
vanishes for SD field strength (\smash{$F^{\a\b} = 0$}).
Hence the single copy and double copy solutions of
$\ell^{\da\a} \propto \tdo^\da\hem \eta^\a$
together defines a solution in Einstein-Maxwell theory:
a SD Taub-NUT instanton
endowed with SD electromagnetic charge.
By using such Einstein-Maxwell configurations
as building blocks,
a generalization of the nonlinear superposition theorem
given in the supplemental material
derives
the Kerr-Newman solution,
stipulating that the KS potential exactly needs to be modified as
\begin{align}
	\label{eq:KN-potential}
	\frac{M}{|\vex{\,-\,}\vex_1|} + \frac{M}{|\vex{\,-\,}\vex_2|}
	- \frac{Q^2}{ |\vex{\,-\,}\vex_1| |\vex{\,-\,}\vex_2| }
	\,.
\end{align}
This reproduces
the complexification prescription
posited in
\rcite{Newman:1965my-kerrmetric}.
The $Q^2$ term in \eqref{eq:KN-potential}
encode
static electromagnetic interaction between
SD and ASD dyons.
Working in the similar fashion,
one derives
the five-parameter subclass of
Pleba\'nski-Demia\'nski \cite{PlebanskiDemianski:1976gy,debever1971type} solutions
with mass, magnetic mass, ring radius, electric charge, and magnetic charge,
in the double KS form.

\skip
\paragraph{Summary}%
This paper elevates the NJA to a rigorous derivation of rotating black hole solutions
with a definite origin:
factorizations into SD and ASD Taub-NUT instantons
and chiral dyons.

This resolution of an old mystery
will spark new perspectives on
modern theoretical studies of rotating black holes:
hidden symmetries \cite{Carter:1968ks,Penrose:1973naked,Compere:2023alp,Guevara:2023wlr}
\footnote{
	The Killing-Yano tensor of Kerr
	decomposes into
	the conformal Killing-Yano tensors
	of the instantons.
},
perturbation theory
\cite{Teukolsky:1973ha,Adamo:2023fbj},
scattering amplitudes
\cite{ahh2017,Guevara:2018wpp,Guevara:2019fsj,chkl2019,aho2020,Johansson:2019dnu,Aoude:2020onz,Lazopoulos:2021mna,zihan2023,fabian2},
and
effective dynamics
\cite{goldberger2006effective,porto2016effective,kalin2020post,vines2018scattering,Guevara:2018wpp,Guevara:2019fsj,chkl2019,Gibbons:1986df,atiyah2014geometry,ferrell1987slow}.

\medskip
\paragraph{Acknowledgements}%
The author thanks
Clifford Cheung,
Thibault Damour,
Maciej Dunajski,
Alfredo Guevara,
Sangmin Lee,
Lionel Mason,
Vinicius Nevoa,
Donal O'Connell,
Julio Parra-Martinez,
Saul Teukolsky,
and
Justin Vines
for discussions or comments.
\JHK is supported by the Department of Energy (Grant No. DE-SC0011632) and by the Walter Burke Institute for Theoretical Physics.
\JHK was also supported 
by Ilju Academy and Culture Foundation.
The author is grateful to
the Galileo Galilei Institute for Theoretical Physics, Florence
and the Isaac Newton Institute for Mathematical Sciences, Cambridge
for hospitality and stimulating environment
during the conference ``The Mathematics behind Scattering Amplitudes'' held in August 2024
and the program ``Twistors in Geometry \& Physics'' held in September 2024,
respectively.
	
	\bibliography{references.bib}
	
	\pagebreak
	\onecolumngrid
\renewcommand{\theequation}{A.\arabic{equation}}
\setcounter{equation}{0}

\begin{center}
	\scshape
	i.\:\:
		Nonlinear Superposition Theorem
\end{center}

The idea of patching SD and ASD spacetimes together
has been a long-standing aspiration
\cite{plebanski1998linear,robinson-TN44-03,robinson1987some}.
Here,
we tackle this problem
in a narrow setup:
\textit{stationary KS spacetimes}.

The complexified Minkowski space $\mflat$
is the linear space $\C^4$ equipped with a holomorphic flat metric $\eta$
with a signature-$(-,+,+,+)$ real section.
Let $x^\m$ be Cartesian coordinates.
Let $u^\m$ be a unit vector such that $u^2 = \eta_{\m\n}\mem u^\m u^\n = -1$;
this is the characterization of the stationary direction $u^\m = \delta^\m{}_0$ in the main article.
All indices are raised and lowered with the flat metric $\eta$.

\skip
\skip
\noindent{\scshape \bfseries Definition 1.1}|%
A null vector field (NVF) $\ell$ on $\mflat$
is a vector field $\ell \in \Gamma(T\mflat)$
such that $\ell^2 = \eta_{\m\n}\mem \ell^\m \ell^\n = 0$.

\skip
\noindent{\scshape \bfseries Definition 1.2}|%
A NVF $\ell$ on $\mflat$
is \textbf{canonical} iff it is
stationary, geodesic, shear-free, 
and unit-normalized as $-u\mdot\ell = 1$.

\skip
\noindent{\scshape \bfseries Proposition 1.3}|%
The first-order derivative $\partial_\m \ell_\n$ 
of a canonical NVF $\ell$
takes the following general form:
\begin{align}
	\label{eq:canonical}
	\partial_\m \ell_\n
	\,=\,
	\frac{\theta}{2}\, 
	\Big(\,{
		\eta_{\m\n}
		+ \ell_\wrap{(\m} (2u{\,-\,}\ell)_\wrap{\n)}
	}\mem\Big)
	+ \frac{\psi}{2}\, \ve_{\m\n\r\s}\hem u^\r\hem \ell^\s
	\,.
\end{align}
Here, the scalar fields $\theta$ and $\psi$ are referred to as 
the \textbf{expansion} and \textbf{twist} of $\ell$,
defined as
$\theta = \partial_\m \ell^\m$
and 
$\psi = 
	u_\m\mem
	\ve^{\m\n\r\s}
	\ell_\n\mem \partial_\r \ell_\s
$.
The \textbf{complex expansions} of $\ell$ refer to
$\rho^\pm = (\theta \pm i\psi)/2$.

(Proposition 1.3)
is easily verified.
For instance, the shear-free property
arises from the fact that
$\eta_{\m\n}
+ \ell_\m u_\n
+ u_\m \ell_\n
- \ell_\m \ell_\n
$ and
$\ve_{\m\n\r\s}\hem u^\r \ell^\s$
in \eqref{eq:canonical}
are tensors lying within the two-dimensional plane
transverse to both $u$ and $\ell$,
spanning the trace-only and rotation modes.
\rcite{vines2018scattering}
makes a fruitful use of the formula in \eqref{eq:canonical}.

\skip
\noindent{\scshape \bfseries Proposition 1.4}|%
From \eqref{eq:canonical}
and its integrability $\partial_\m(\partial_\n\ell_\r) = \partial_\n(\partial_\m\ell_\r)$, it follows that
$\ell^\m\mem \partial_\m \rho^\pm = - (\rho^\pm)^2$,
$\ell^\m\mem \ell^\n\mem \partial_\m \partial_\n \rho^\pm = 2\hem (\rho^\pm)^3$,
and
$\partial^2 \rho^\pm = 0$.
The first equation is
the Sachs equation \cite{penrose1984spinors2,huggett1994introduction} in flat background.

\skip
\skip
\noindent{\scshape \bfseries Definition 2.1}|%
For each canonical NVF $\ell$ on $\mflat$,
the \textbf{single copy solution} is the abelian gauge connection $A_\m = \theta\mem \ell_\m$
while the \textbf{double copy solution} is the metric $g_{\m\n} = \eta_{\m\n} + \e\, \theta\mem \ell_\m \ell_\n$,
where $\e$ is a constant parameter.

\skip
\noindent{\scshape \bfseries Proposition 2.2}|%
The single copy and double copy solutions of a canonical NVF
solve vacuum flat-background Maxwell's equations and vacuum Einstein's equations
in the holomorphic category,
respectively.

\skip
\noindent{\scshape \bfseries Definition 2.3}|%
A canonical NVF $\ell$ is \textbf{self-dual} (SD) iff
its expansion is nonvanishing 
and
the field strength $F_{\m\n} = \partial_\m (\theta\mem \ell_\n) - \partial_\n (\theta\mem \ell_\m)$
of its single copy solution is SD.
A canonical NVF $\ell$ is \textbf{anti-self-dual} (ASD) iff
its expansion is nonvanishing 
and
its field strength is ASD.

\skip
\noindent{\scshape \bfseries Proposition 2.4}|%
A canonical NVF is SD iff
$\rho^+ \neq 0$ and $\rho^- = 0$.
A canonical NVF is ASD iff
$\rho^+ = 0$ and $\rho^- \neq 0$.

\skip
\noindent{\scshape \bfseries Proposition 2.5}|%
The double copy solution of a SD (ASD) canonical NVF
exhibits SD (ASD) curvature:
${*}R_{\m\n\r\s} = \pm i\mem R_{\m\n\r\s}$.

(Proposition 2.5) can be shown by
employing the orthonormal coframe
$e^A{}_\m = \delta^A{}_\m\mem (\delta^\m{}_\n + \e\, \theta\mem \ell^\m \ell_\n/2)$,
in which case the spin connection is
\begin{align}
	\label{sc}
	\gamma_{AB\r}
	\,=\,
	- \frac{\e}{2}\,
		F_{\m\n}\mem \delta^\m{}_A\mem \delta^\n{}_B\,
		\ell_\r
	- \frac{\e}{2}\,
	\BB{
		\rho^+\mem \Pi^+_{ABCD}
		+
		\rho^-\mem \Pi^-_{ABCD}
		\nem}\mem 
	\BB{
		\delta^C{}_\r
		+ \delta^C{}_\k\mem u^\k \ell_\r
	}
	\BB{
		\delta^D{}_\s\mem A^\s
	}
	\,,
\end{align}
where the SD and ASD projectors are
$\Pi^\pm_{ABCD} = \frac{1}{2}\mem(
	\eta_{AC} \eta_{BD}
	+ \eta_{AD} \eta_{BCD}
	- i\mem \ve_{ABCD}
)$.

\skip
\skip
\noindent{\scshape \bfseries Lemma 3.1} (\textit{Little Superposition Lemma})|%
Suppose a collection of canonical NVFs
$\ell_i$ ($i=1,2,\cdots n$)
such that 
$\sum_i c_i\mem \theta_i = 0$
for a set of constants $c_i$.
If there exists a scalar function $\chi$ such that
$\sum_i c_i\, \theta_i \ell_i{}_\m = \partial_\m \chi$,
then there exists a one-form field $\xi_\m$ such that
$\sum_i c_i\, \theta_i \ell_i{}_\m \ell_i{}_\n = 2\mem \partial_\wrap{(\m} \xi_\wrap{\n)}$.
Explicitly, $\xi_\m = \chi\mem u_\m - \protect\sum_i c_i\, \ell_i{}_\m$.

\skip
\noindent{\scshape \bfseries Lemma 3.2} (\textit{Superposition Lemma})|%
If $\ell_{1,2}$ are canonical NVFs
with complex expansions $\rho_{1,2}^\pm$,
then
\begin{align}
	\label{eq:elformulas}
	\ell^{\da\a}
	\,=\, \frac{
		\ell_1^{\da\b} \delta_{\b\db}\mem \ell_2^{\db\a}
	}{1{\,+\,}\ell_1\nem\mdot\ell_2/2}
	\,,\quad
	\tell^{\da\a}
	\,=\, \frac{
		\ell_2^{\da\b} \delta_{\b\db}\mem \ell_1^{\db\a}
	}{1{\,+\,}\ell_1\nem\mdot\ell_2/2}
\end{align}
are canonical NVFs with respective complex expansions
$(\rho^+,\rho^-) = (\rho^+_1,\rho^-_2)$
and
$(\trho^+,\trho^-) = (\trho^+_2,\trho^-_1)$,
satisfying
\begin{align}
	\label{eq:chi12}
	\theta\mem \ell_\m
	+ \ttheta\mem \tell_\m
	\,=\,
	\theta_1\mem \ell_1{}_\m
	+ \theta_2\mem \ell_2{}_\m
	+ \partial_\m\chi
	\qiq
	\theta\mem \ell_\m \ell_\n
	+ \ttheta\mem \tell_\m \tell_\n
	\,=\,
	\theta_1\mem \ell_1{}_\m \ell_1{}_\n
	+ \theta_2\mem \ell_2{}_\m \ell_2{}_\n
	+ 2\mem \partial_\wrap{(\m} \xi_\wrap{\n)}
\end{align}
with $\chi = \log( (1 {\mem+\,} \ell_1\nem\mdot\ell_2/2)^2 )$
and 
$\xi_\m = \chi\mem u_\m - \ell_\m - \tell_\m + \ell_1{}_\m + \ell_2{}_\m$.
Here, \smash{$\delta_{\a\da} = -u_\m (\s^\m)_{\a\da}$}.

The geodesic shear-free condition plays an important role in (Lemma 3.1).
(Lemma 3.2) follows by direct computation
and (Lemma 3.1),
where the chiral nature of \eqref{eq:elformulas} is crucial.

\skip
\noindent{\scshape \bfseries Theorem 3.3} (\textit{Nonlinear Superposition Theorem})|%
A SD canonical NVF $\ell_1$
and an ASD canonical NVF $\ell_2$
together defines
an exact solution 
$g_{\m\n} = \eta_{\m\n} + \e\mem \big(\mem{ \theta_1 + \theta_2 }\mem\big)\mem \ell_\m\ell_\n$
to Einstein's equations
via 
\smash{$
\ell^{\da\a}
=
	\ell_1^{\da\b} \delta_\wrap{\b\db}\mem \ell_2^{\db\a}
	/({1{\,+\,}\ell_1\nem\mdot\ell_2/2})
$}.

Proof is straightforward from (Lemma 3.2)
and (Definition 2.3):
$(\trho^+,\trho^-) = (\trho^+_2,\trho^-_1)$ vanishes.
This describes the nonlinear superposition of SD and ASD vacuum metrics in the sense that
the Riemann tensor of
$\eta_{\m\n} + \e\, \theta\mem \ell_\m \ell_\n$
equals the sum of the Riemann tensors of
$\eta_{\m\n} + \e\, \theta_1\mem \ell_1{}_\m \ell_1{}_\n$
and
$\eta_{\m\n} + \e\, \theta_2\mem \ell_2{}_\m \ell_2{}_\n$
at leading order in the parameter $\e$.

\newpage
\noindent{\scshape \bfseries Lemma 4.1} (\textit{Maxwell Stress-Energy via Modifying the KS Potential})|%
If $\ell$ is a canonical NVF on $\mflat$
with complex expansions $\rho^\pm$,
then
the gauge potential
$A_\m = \theta\mem \ell_\m$
and 
the metric
\smash{$g_{\m\n} = \eta_{\m\n} + \e\mem \big(\mem{ \theta -2\alpha \mem \rho^+ \rho^- }\mem\big)\mem \ell_\m\ell_\n$}
together defines a solution to 
$G^\m{}_\n[g] = \e\mem\alpha\, T^\m{}_\n[A,g]$
and $\nabla^\n F_{\m\n} = 0$:
Einstein-Maxwell theory
in the holomorphic category.
Here,
$G^\m{}_\n[g]$ is the Einstein tensor associated with the metric $g$,
$T^\m{}_\n[A,g]$ is the Maxwell stress-energy tensor,
$\nabla$ is the Levi-Civita connection of the metric $g$,
$F_{\m\n} = 2\mem \nabla_\wrap{[\m} A_\wrap{\n]} = 2\mem \partial_\wrap{[\m} A_\wrap{\n]}$
is the field strength,
and $\a$ is a constant parameter.

\skip
\noindent{\scshape \bfseries Theorem 4.2} (\textit{Nonlinear Superposition for Einstein-Maxwell Solutions})|%
A SD canonical NVF $\ell_1$
and an ASD canonical NVF $\ell_2$
together defines an exact solution
$A_\m = \big(\mem{ \theta_1 + \theta_2 }\mem\big)\mem \ell_\m$,
$g_{\m\n} = \eta_{\m\n} + \e\mem \big(\mem{ \theta_1 + \theta_2 -2\alpha \mem \theta_1 \theta_2 }\mem\big)\mem \ell_\m\ell_\n$
to Einstein-Maxwell theory
via 
\smash{$
	\ell^{\da\a}
	=
	\ell_1^{\da\b} \delta_\wrap{\b\db}\mem \ell_2^{\db\a}
	/({1{\,+\,}\ell_1\nem\mdot\ell_2/2})
$}.
Here, $Q$ is a constant parameter such that
$\e\hem\alpha = Q^2/2$.

For deriving the Kerr-Newman solution,
one rescales the gauge potential
by $Q/8\pi\varepsilon_0$
so that the electromagnetic charges of the SD and ASD solutions
become $(Q/2, \mp\hem i\hem Q/2)$.
Consequently,
the constant parameters are taken as
$\e = GM/c^2$
and 
$\e\hem \alpha 
= (8\pi G/c^4) \mdot (1/\mu_0) \mdot (Q/8\pi\ve_0)^2 / c^2
= Gk_\text{e} Q^2 \nem/2c^4
$.
Hence,
in the relativists' natural unit
($c{\,=\,}1$, $G{\,=\,}1$, $k_\text{e}{\,=\,}1$),
one finds
$\e = M$ and $\e\hem \alpha = Q^2 \hnem/2$
as stated in the main article.

\skip
\skip
\noindent{\scshape \bfseries Theorem 5.1} (\textit{Effective Linearization})|%
A metric of the form $g_{\m\n} = \eta_{\m\n} + \phi\mem \ell_\m\hem \ell_\n$
with $\ell^2 = 0$
solves
the exact vacuum Einstein's equations
if it solves
the linearized vacuum Einstein's equations
\cite{xanthopoulos1978exact,Harte:2016vwo,vines2018scattering}.

\skip\skip\skip\skip
\begin{center}
	\scshape
	ii.\:\:
	NJA in Spheroidal Coordinates
\end{center}

The main article describes the inverse metrics in 
\eqrefs{eq:ig-kerr}{eq:ig-sch}
in the spinor language.
The original paper \cite{Newman:1965tw-janis},
on the other hand,
uses
the null tetrad language
in spheroidal coordinates.
We show that they are equivalent.

\skip\skip\noindent
\textbf{Null Tetrad}.|%
Given two spinors
$\tdo_1^\da$ and $o_2^\a$
and $\delta_{\a\da} = -u_\m\mem (\s^\m)_{\a\da}$,
a null tetrad in flat spacetime is formed by
\begin{align}
	\label{eq:tetrad-flat}
	\ell^{\da\a} =
	\frac{\tdo^\da_1 o^\a_2}{ 
		\langle o_2 | \delta | \tdo_1 \rsq
	}
	\,,\,\,\,
	n^\circ{}^{\da\a} =
	\frac{\ti^\da_2 \iota^\a_1}{ 
		\langle o_2 | \delta | \tdo_1 \rsq
	}
	\,,\,\,\,
	m^{\da\a} =
	\frac{\tdo^\da_1 \iota^\a_1}{ 
		\langle o_2 | \delta | \tdo_1 \rsq
	}
	\,,\,\,\,
	\tm^{\da\a} =
	\frac{\ti^\da_2 o^\a_2}{ 
		\langle o_2 | \delta | \tdo_1 \rsq
	}
	\,\,\,\implies\,\,\,
	\eta^{\m\n} = -\ell^{(\m} n^\circ{}^{\n)} + m^{(\m} \tm^{\n)}
	\,,
\end{align}
where
$\ti_2{}_\da = - o^\a_2\mem \delta_{\a\da}$
and
$\iota_1{}_\a = - \delta_{\a\da}\mem \tdo^\da_1$
so that
$\lsq \tdo_1 \ti_2 \rsq = \langle o_2 | \delta | \tdo_1 \rsq$
and
$\langle \iota_1 o_2 \rangle = \langle o_2 | \delta | \tdo_1 \rsq$.

For example,
the principal spinors of the Schwarzschild solution are given by
$\tdo^\da = (1,(x{\,+\,}iy)/(|\vex|{\,+\,}z))$
and
$o^\a = (1,(x{\,-\,}iy)/(|\vex|{\,+\,}z))$.
Taking
$\tdo^\da_1 = \tdo^\da$
and
$o^\a_2 = o^\a$
in \eqref{eq:tetrad-flat}
gives
$\ell^\m = (1,x/|\vex|,y/|\vex|,z/|\vex|)$,
which is the outgoing canonical NVF 
for the KS metric of the Schwarzschild solution.

The physical spinor of SD/ASD Taub-NUT instanton
equals the SD/ASD principal spinor of Schwarzschild
\cite{note-sdtn}.

The main article sets
$\tdo_1^\da$ and $o_2^\a$
as the physical spinors of the SD and ASD Taub-NUT instantons
at $\vex_1 = +i\vea$ and $\vex_2 = -i\vea$
with $\vea = (0,0,a)$.
Explicitly, this means to take
\smash{$\tdo^\da_1 = (1,\tzeta)$},
\smash{$o^\a_2 = (1,\zeta)$},
\smash{$\ti^\da_2 = (-\tzeta,1)$},
\smash{$\iota^\a_1 = (-\zeta,1)$},
and
$1/\langle o_2|\delta|\tdo_1\rsq = (\robl{\,+\,}z)/2\robl$,
where
\begin{align}
	\tzeta
	\,=\, \frac{x+iy}{|\vex{\,-\,}i\vea|+(z{\,-\,}ia)}
	\,=\, \frac{x+iy}{\robl+z}\mem \frac{\robl}{\robl-ia}
	\,,\quad
	\zeta
	\,=\, \frac{x-iy}{|\vex{\,+\,}i\vea|+(z{\,+\,}ia)}
	\,=\, \frac{x-iy}{\robl+z}\mem \frac{\robl}{\robl+ia}
	\,.
\end{align}
Here, $\robl$ denotes the function of $x,y,z$ defined in \eqref{eq:r-oblate}.
A branch cut prescription
\cite{Newman:2002mk,aho2020}
sets $|\vex{\,\mp\,}i\vea| = \sqrt{(\vex{\,\mp\,}i\vea)^2} = \robl \mp iaz/\robl$.
The computation of the resulting flat null tetrad due to \eqref{eq:tetrad-flat}
is straightforward.

\skip\skip\noindent
\textbf{Spheroidal Coordinates}.|%
A nice choice of angle variables $(\theta,\varphi)$
simplifies the result as
$\ell^\m = (1, \sin\theta \cos\varphi,$ $ \sin\theta\sin\varphi, \cos\theta)$:
$x \pm iy = (\robl \mp ia)\mem \sin\theta\mem e^{\pm i\varphi}$,
$z = \robl\mem \cos\theta$.
In fact, $(r,\theta,\varphi)$ is known as
the twisted oblate spheroidal coordinate system.
As a result,
the components of $\ell$ are simply $(1,1,0,0)$
in the $(t,\robl,\theta,\varphi)$ coordinate system.
Thus, by introducing the retarded time $u = t-\robl$,
the null tetrad is
\begin{align}
	\label{eq:njatetrad-flat}
	\ell \,=\, \partial_\robl
	\,,\quad
	n^\circ \,=\, 2\partial_u - \partial_\robl
	\,,\quad
	m \,=\, \frac{e^{i\varphi}}{\robl{\,+\,}ia\cos\theta}\mem
	\bigg(\,{
		ia\sin\theta\mem \Big(\mem{
			\partial_u - \partial_\robl
		}\mem\Big)
		+ \partial_\theta
		+ \frac{i}{\sin\theta}\mem \partial_\varphi
	}\,\bigg)
	\,,\quad
	\tm \,=\, 
		m \mem\big|_{i \to -i}
	\,.
\end{align}

The null tetrad $(\ell,n,m,\tm)$ for
the curved inverse metric in \eqref{eq:ig-kerr} 
follows by
perturbing $n^\circ \to n = n^\circ + M\mem \theta\mem \ell$:
\begin{align}
	\label{eq:njatetrad-n}
	n
	\,=\,
	2\partial_u - \partial_\robl
	+ M\mem 
	\bigg({
		\frac{1}{\robl{\,-\,}ia\cos\theta}
		{\,+\,} 
		\frac{1}{\robl{\,+\,}ia\cos\theta}
	}\bigg)
	\, \partial_\robl
	\,.
\end{align}
\eqrefs{eq:njatetrad-flat}{eq:njatetrad-n}
are literally Eq.\,(7) of \rcite{Newman:1965tw-janis}.
Taking the limit $a\to0$ exactly reproduces Eq.\,(4) of \rcite{Newman:1965tw-janis}.

The NJA inverts this limit.
\rcite{Newman:1965tw-janis} views it as
a ``complex coordinate transformation''
due to the suggestiveness of the spheroidal coordinates.
However, the main article clarifies that
the NJA is not a coordinate transformation.

By repeating the above exercise,
one systematically derives the NJA for
Kerr-Newman, $\text{Kerr}^\star$, Kerr-Taub-NUT, and charged Kerr-Taub-NUT
(five-parameter Pleba\'nski-Demia\'nski)
solutions.
We emphasize that
what facilitates such a systematic and universal derivation of NJAs
is the very instanton pair structure
elucidated in the main article.
In other words,
the origin of the NJA is the factorization of black holes into Taub-NUT instantons and chiral dyons.

Note that this derivation of the NJA works in any signature:
Euclidean, Lorentzian, Kleinian, mostly-plus, mostly-minus, etc.
For instance,
one can employ twisted prolate hyperboloidal coordinate system
$x \pm iy = (\rprol \pm ja)\mem \sin\theta\mem e^{\pm j\varphi}$,
$z = \rprol \cos\theta$,
where $j$ denotes the unit split complex number ($j^2 = 1$).
	
\end{document}